\documentclass[aps,pre,showpacs,superscriptaddress,twocolumn]{revtex4}
\usepackage{amsfonts}
\usepackage[dvips]{graphicx}

\usepackage{hyperref} % para hipervincular las referencias en el documento
\usepackage{epsfig}
\usepackage{color}
\usepackage{mathrsfs} 

\usepackage{syntonly}
\usepackage{float} % Para ubicaion de imagenes flotantes
\usepackage{amsthm,graphics,amssymb,amsmath,latexsym}
\usepackage{fancyhdr}
\newcommand{\mre}{\mathrm{e}}
\newcommand{\mrd}{\mathrm{d}}
\newcommand{\order}[1]{\mathcal{O}\left(#1\right)}
\begin{document}

\title{Finite-size corrections to scaling of the magnetization distribution \\
in the $2d$ $XY$-model at zero temperature}
\author{G. Palma}
\email{guillermo.palma@usach.cl}
\affiliation{Departamento de F\'{i}�sica, Universidad de Santiago de Chile,
Casilla 307, Santiago 2, Chile}

\author{F. Niedermayer}
\affiliation{Albert Einstein Center for Fundamental Physics, Institute for Theoretical Physics,\\ 
University of Bern, Switzerland}

\author{Z. R\'{a}cz}
\affiliation{MTA-ELTE Theoretical Physics Research Group, 
Budapest, Hungary}

\author{A. Riveros}
\affiliation{Departamento de F\'{i}�sica, Universidad de Santiago de Chile,
Casilla 307, Santiago 2, Chile}

\author{D. Zambrano}
\affiliation{Departamento de F\'{i}�sica, Universidad T\'{e}cnica Federico Santa Mar\'{i}�a, 
Av. Espa\~{n}a 1680, Casilla 110-V, Valpara\'{\i}so, Chile}

\begin{abstract}

  The zero-temperature, classical $XY$-model on an $L \times L$ square-lattice
  is studied by exploring the distribution $\Phi_L(y)$ of its centered and
  normalized magnetization $y$ in the large $L$ limit.  An integral 
  representation of the cumulant generating function, 
  known from earlier works, is used for the numerical evaluation of
  $\Phi_L(y)$, and the limit distribution $\Phi_{L 
    \rightarrow \infty} (y) = \Phi_0(y)$ is obtained with high precision. The
  two leading finite-size corrections $\Phi_L (y) -\Phi_0 (y) \approx a_1(L)\,
  \Phi_1(y) + a_2(L)\,\Phi_2(y)$ 
  are also extracted both from numerics and from analytic calculations. 
  We find that the amplitude $a_1(L)$ scales as $\ln(L/L_0) /L^2$ and the
  shape correction function $\Phi_1 (y)$ can be expressed 
  through the low-order derivatives of the limit distribution, $\Phi_1 (y) =
  [\,y\, \Phi_0 (y) + \Phi'_0 (y)\,]'$. The second finite-size correction  
  has an amplitude $a_2(L)\propto 1/L^2$ and one finds that 
  $a_2\,\Phi_2(y) \ll a_1 \,\Phi_1(y)$ already for small system size ($L> 10$).
  We illustrate the feasibility of observing  the calculated finite-size 
  corrections by performing simulations of the $XY$-model at low temperatures, 
  including $T = 0$.
 
\end{abstract}

\pacs{05.50.+q, 02.70.-c, 68.35.Rh, 75.40.Cx}

\maketitle

% ****************************************************************************
\section{Introduction}
% ****************************************************************************

Studies of the scaling properties of fluctuations have played an important
role in developing the theory of equilibrium critical phenomena, and they also 
proved to be instrumental in exploring systems driven far from equilibrium 
where fluctuations often diverge in the thermodynamic limit. In particular, 
finding critical exponents through finite-size (FS) scaling has become a standard 
method for establishing universality classes \cite{Fisher}-\cite{FSS_Num_stat} 
and, furthermore, the shapes of the distribution functions of fluctuations have
also been used as hallmarks of universality classes for both equilibrium
\cite{Bruce}-\cite{Ex_wide_applic} and non-equilibrium systems
\cite{Racz_et_al}-\cite{Korniss_et_al}.

One of the most studied distribution of critical order-parameter fluctuations
is that of the two dimensional classical $XY$ model. The model is
important in itself since it serves as a prime example of an equilibrium phase
transition with topological order \cite{Kosterlitz_Thouless, Chaikin}. Recent 
interest comes also from a suggestion that its critical
magnetization distribution, $P_L(m)$, may describe the fluctuations in a
remarkable 
number of diverse far-from-equilibrium steady states. Examples range
from the energy-dissipation in turbulence \cite{Bramwell_nature,Aji_Goldenfeld} 
and interface fluctuations in surface evolution 
\cite{Racz_et_al,Korniss_et_al} to electroconvection in liquid crystals
\cite{Toth_prl}, as well as to river-height fluctuations
\cite{Bramwell_epl}. In some of the examples, such as the surface growth
problems described by the Edwards-Wilkinson model \cite{Edwards_Wilkinson},
one can establish a rigorous link to the low-temperature limit 
(spin-wave approximation) of the $XY$ model. Furthermore, the non-gaussian
features  
of $P_L(m)$ such as its exponential and double exponential 
asymptotes appear to have generic origins
\cite{Bramwell_PRE2001,Bramwell_nat_phys2009}, thus 
explaining the observed collapse in a remarkable set of experimental and
simulation data.  
In the majority of the examples, however, the experiments or
simulations provide data of insufficient accuracy to decide unambiguously
about the universality class. Indeed, it is not easy even to observe
the changes  
occurring in $P_L(m)$ if the $XY$ model is considered at finite temperatures where 
the spin-wave approximation breaks down \cite{Palma_2005,Palma_2006,Banks_2005}. 

In order to be more confident about universality claims, one would like to be
able  
to examine the distribution functions in more details. A well known and much 
investigated method of extracting additional information in critical systems  
such as the zero-temperature $XY$ model is the study of the FS behavior 
\cite{Fisher}-\cite{FSS_Num_stat}. The implication of FS scaling  
is that the appropriately scaled distribution function has a well defined limit 
when the system size tends to infinity ($L\to \infty$). A frequently used
scaling variable is 
the centered and normalized magnetization $y=(m-\langle m\rangle)/\sigma_m$
where  
$\sigma_m^2=\langle m^2\rangle-\langle m\rangle^2$. This choice of scaling
variable  
eliminates possible divergences in $\langle m\rangle$ and, in general,
produces a non-degenerate limit distribution $\Phi_0(y)$ \cite{Antal_et_al_prl}:
\begin{equation} \label{limdist}
  \lim\limits_{L\to\infty} \Phi_L(y) 
  \equiv\lim\limits_{L\to\infty}\sigma_m \, P_L( \langle m\rangle+\sigma_m \, y )  
  = \Phi_0(y)\, .
\end{equation}
The function $\Phi_L(y)$ can be expressed in a compact form for the
zero-temperature $XY$ model  
\cite{Archam_1998}. Its large-$L$ limit, $\Phi_0(y)$,
has been numerically evaluated and compared with simulations and experiments
on a variety of systems  
(see e.g. \cite{Bramwell_nature,Bramwell_PRE2001,Bramwell_epl}).

Once the limit distribution is known, one can investigate the approach 
to $\Phi_0(y)$ as the system size increases. As we shall show below, keeping the
leading and the next to leading terms, the FS corrections, 
$\delta \Phi_L(y) = \Phi_L(y) - \Phi_0(y)$,
to the limit distribution can be written as
\begin{equation}  \label{lead-c1}
\delta \Phi_L(y) 
  = a_1(L)\,\Phi_1(y) + a_2(L)\,\Phi_2(y) + \order{\ln^2\hspace{-0.1cm}L/L^4} \,.
\end{equation}
We kept two terms since the asymptotic $L$-dependence of the 
leading term differs from the next one only by a slowly varying logarithmic factor.
Indeed, our calculation yields the amplitudes $a_i(L)$ in the following form
\begin{equation} \label{a12}
  a_1(L) = \frac{\alpha \ln L + \gamma}{L^2} 
  =  \frac{\alpha \ln (L/L_0)}{L^2}\,,
  \hspace{0.4cm} a_2(L)=\frac{\gamma'}{L^2} \,.
\end{equation}
Here, the coefficient $\alpha$ of the logarithmic term can be expressed through 
the Catalan constant $G$ as $\alpha = 3\pi/4G =  2.572...\,$. 
The coefficients of the $1/L^2$ terms are 
$\gamma=-2.803$ (giving $L_0=2.973$) and $\gamma'= 2\pi/3$ 
(for details, see Sec.\ref{FSS_section}).

A remarkable result about the function $\Phi_1(y)$ characterizing 
the leading shape correction to the limit distribution is that 
it can be expressed through $\Phi_0(y)$ as
\begin{equation}\label{Phi_1}
  \Phi_1(y) = \left[ \,y \,\Phi_0(y) + \Phi'_0(y)\,\right]' \,,
\end{equation}
where the prime denotes the derivative over $y$. The second scaling function 
$\Phi_2(y)$ is more complex
in the sense that it cannot be expressed through a finite sum 
of the derivatives of the limit distribution. 
However, it can be calculated efficiently using integral 
representations as explained in Sec.\ref{FSS_section} [see Eq.\eqref{Phi12}] and in 
Appendix \ref{Appendix_B} [see Eq.\eqref{Psi2_int}].

Evaluating the scaling functions $\Phi_i(y)$ numerically, one observes that 
$a_1(L)\Phi_1(y) \gg a_2(L)\Phi_2(y)$ already for small systems ($L> 10$).
This leads to one of the main conclusions of our work, namely that 
the FS corrections to the limit distribution can be written to an excellent 
accuracy in the following form
\begin{equation} \label{nextlead-c2}
    \delta \Phi_L(y) \approx  
     \alpha\frac{\ln (L/L_0)}{L^2}\,\left[ \,y\, \Phi_0 (y) 
      + \Phi^\prime_0 (y) \,\right]^\prime\,. 
\end{equation}
The above expression can be easily calculated and compared with experiments 
and simulations using both the scaling of the amplitude and checking the shape of 
the correction. 

It is important to note that $\Phi_1(y)$ is uniquely determined by 
the limit distribution, thus the leading shape correction 
has the same universality attributes as 
the limit distribution itself. 
In renormalization group language, the meaning of the above result is that 
the eigenfunction corresponding to the direction of slowest 
approach to the fixed point distribution can be expressed through 
the limit distribution. A simple and transparent derivation of an analogous 
result for the case of the central limit theorem can be found in \cite{Jona}. 

In order to show the workings of the method of FS scaling, 
we carried out Monte Carlo (MC) simulations of the $XY$ model in its low-temperature limit, 
including its $T = 0$-limit. We examined the FS corrections
described above for the lattice of size $L = 10$, as an illustration, and found close agreement
between the theoretical and MC results.

The paper is organized as follows. The $XY$ model and the zero-temperature
magnetization distribution is described in Sec.\ref{model}. Next, the FS corrections 
are calculated in Sec.\ref{FSS_section} with 
the technicalities relegated to two Appendices (the direct calculations of the
needed cumulants are found in Appendix \ref{Appendix_A}, while an integral
representation that simplifies the evaluation of  
finite-volume sums is presented in Appendix \ref{Appendix_B}). 
Finally in Sec.\ref{MC_sect}, and as an illustration, we have compared the FS results 
with the MC simulations for a lattice of size $L = 10$.

% *****************************************************************************
\section{Probability density of the magnetization for the classical $XY$-model in two-dimensions} 
\label{model}
% *****************************************************************************

The two-dimensional, classical $XY$ model on a square lattice 
is defined by the Hamiltonian
\begin{equation} \label{XYham}
H=-J\sum\limits_{\langle i,j\rangle}\cos(\theta_i-\theta_j)
\end{equation}
where the angle variables $\theta_i$ describe the orientation of unit vectors in 
the plane, $J>0$ is the ferromagnetic interaction strength between nearest neighbor 
vectors on a periodic, $La\times La$ square lattice. From now on we will set $J=1$ and $a=1$, which defines 
the energy- and length scales in the problem. 

The order parameter $m$ whose probability distribution is of our interest 
is defined as
\begin{equation} \label{m_order}
m=\frac{1}{L^2}\sum\limits_{i} \cos(\theta_i-\bar\theta)
\end{equation}
where $\bar\theta$ is the instantaneous average orientation.
The probability density function (PDF) of the magnetization, $P_L(m)$, has been much 
studied 
\cite{Archam_1998,Bramwell_nature,Bramwell_epl,Bramwell_PRE2001,
Bramwell_nat_phys2009,Palma_2005,Palma_2006,Banks_2005}.
Its zero temperature limit ($T\to 0$) has been first calculated \cite{Archam_1998}
by using the spin-wave approximation 
$\cos(\theta_i-\theta_j)\approx 1-(\theta_i-\theta_j)^2/2$ and summing up the moments series 
for the PDF. 

Another, field-theoretic, approach for obtaining $P_L(m)$ is based on evaluating the 
Fourier transform of the partition function of the system in a loop expansion 
\cite{Palma_2005}. It turns out that the $n$-loops contribution corresponds 
to the $T^{\, n-1}$ contribution, where
$T$ is the system temperature. Correspondingly, the expansion up to one-loop gives exactly 
the zero-temperature limit of $P_L(m)$.

Both of the above methods yield an identical result to leading order. 
The second method, however, shows explicitly how to compute the higher temperature 
corrections to $P_L(m)$ through an effective, $T$-dependent lattice propagator, 
$\Gamma = (I+\frac{ikT}{L^2}\mathcal{G})^{-1} \mathcal{G}$, where $\mathcal{G}$ is 
independent of $T$ \cite{Cardy}, and is defined by its Fourier representation as
\begin{equation}
  \mathcal{G}(\mathbf{k}) = 1/ \, \mathbf{\hat{k}}^2 \hspace{0.2cm}.
\end{equation}
Here, the components of the vector $\mathbf{\hat{k}}$ are
$\hat{k}_i =2 \sin(k_i/2)$, where $k_i = 2 \pi n_i/L$ with 
$i =1,2$ and $n_i \in Z$ such that the lattice momentum $\mathbf{k}$
lies in the first Brillouin zone: 
$- \pi < k_i \leq \pi$.

Hence the PDF at $T = 0$ can be obtained by the well known `$1/2 \, \text{Trace}( \ln )$' expression of the one-loop result \cite{Palma_2005}
\begin{eqnarray} \label{analytic_Pm}
\Phi_L(y) \hspace{-0.2cm} &&= \int_{-\infty}^{\infty} \frac{dq}{2\pi} \hspace{0.1cm} \sqrt{\frac{g_2}{2}} 
 \exp{\left(i\sqrt{\frac{g_2}{2}} q y\right)} \times \hspace{1.9cm}\nonumber \\
  && \exp{\left[- \frac{1}{2} \sum_{\mathbf{k}
        \neq 0} \hspace{-0.1cm} \left\lbrace \ln{\left(1 - \frac{iq}{L^2}
            \mathcal{G}(\mathbf{k})\right)} 
        \hspace{-0.15cm}+ \hspace{-0.05cm}\frac{iq}{L^2} \mathcal{G}(\mathbf{k}) \right\rbrace \right]} 
\end{eqnarray}
where $y$ is the centered and normalized magnetization $y = (m - \langle m
\rangle )/\sigma_m$, and the coefficient $g_2$ is defined as the particular $n = 2$ case of the general 
expression:
\begin{equation} \label{gndef}
  g_n \equiv g_n(L)= \sum_{ \mathbf{k} \neq 0} \mathcal{G}(\mathbf{k} )^n /L^{2n} \,.  
%  \frac{\mathcal{G}(\mathbf{k} )^n }{L^{2n}} \,.  
\end{equation}
The PDF in Eq.\eqref{analytic_Pm} agrees with the corresponding formula 
obtained in \cite{Bramwell_PRE2001}, (see Eq.(26) therein).

\begin{figure}[H]
  \centering \vspace{-0.2cm} \hspace{-0.2cm}
  \includegraphics[width= 1.0\linewidth]{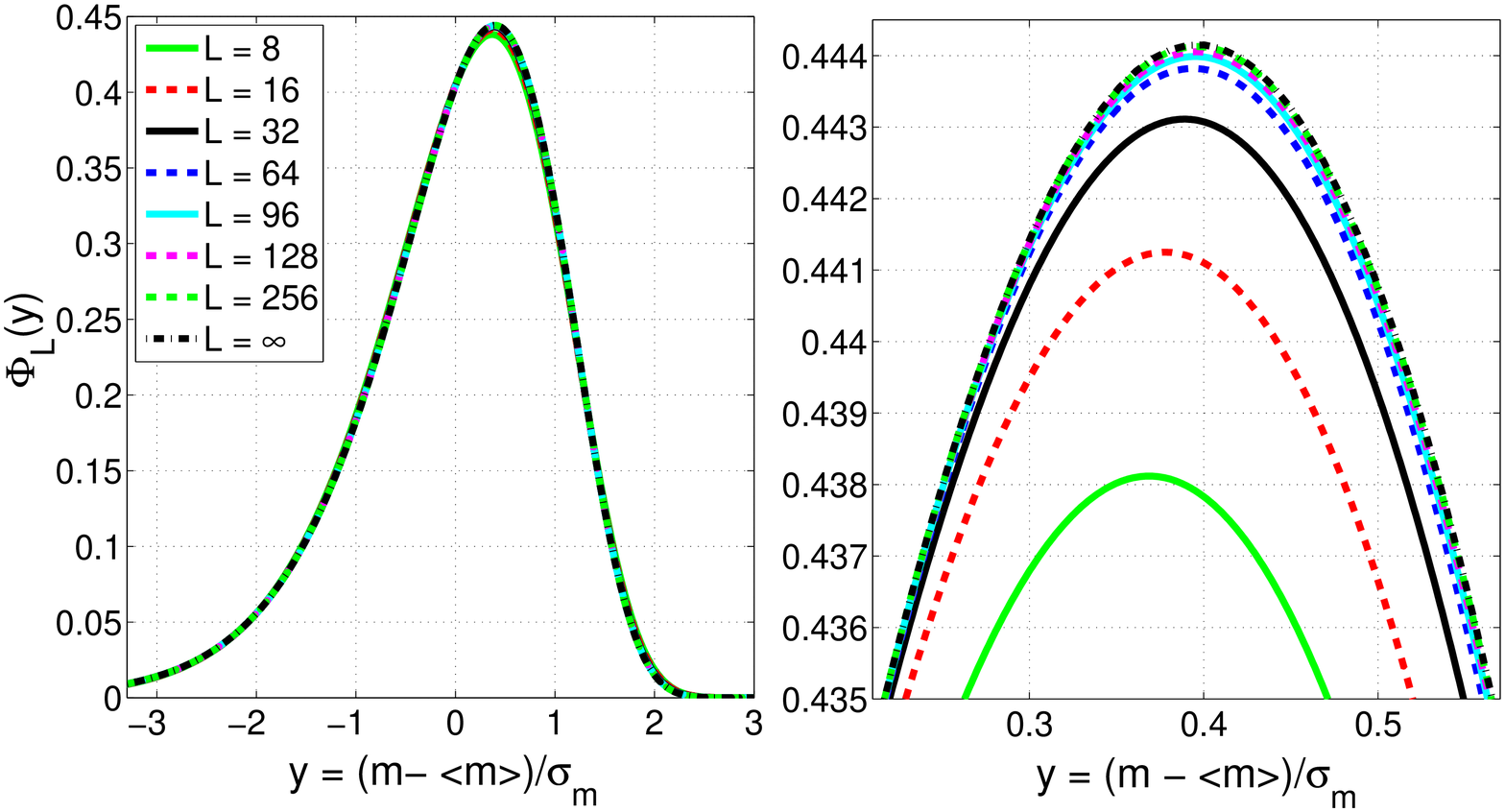}
  \vspace{-0.5cm}
  \caption{\textit{The scaled probability density of the magnetization $\Phi_L (y)$, 
  computed numerically using Eq.\eqref{analytic_Pm},
  is displayed for various lattice sizes $L$. The remarkably fast convergence 
  to the limit distribution $\Phi_0 (y)$ is demonstrated in the left panel 
  while, to illustrate the corrections to the limit distribution, the 
  peak region of the $\Phi_L (y)$ is magnified in the right panel.}}
  \label{PDF_1loop_fig}
\end{figure}

In Fig.\ref{PDF_1loop_fig} we plot $\Phi_L(y)$ as a function of $y$
using Eq.(\ref{analytic_Pm}) for different $L$ values. On the right
panel, we display a zoom into the region close to the peak of the
distribution functions. It can be clearly seen that $\Phi_L(y)$ tends towards
an asymptotic distribution $\Phi_0(y)$ as $L\to \infty$ and, furthermore, 
one observes that the convergence is fast. Nevertheless, in experiments, the value
of $L$ is not known and deviations from $\Phi_0(y)$ are observed. They are often 
explained away as finite-size effects and thus leaving the universality claims 
not entirely justified. 
We would like to emphasize that there is information in the FS corrections 
(shown on Fig.\ref{PDF_1loop_fig} around the peak of the PDF) and, 
by evaluating these corrections, one may refine the reasoning for or against 
finding a universality class.

% *****************************************************************************
\section{Finite-Size Corrections to the Limit Distribution}
\label{FSS_section}
% *****************************************************************************

As explained in Sec.\ref{model}, the PDF of the magnetization for the $2d$
$XY$-model at $T = 0$ is given by the 1-loop analytic
expression of Eq.(\ref{analytic_Pm}), and the numerical evaluation of the 
$L\to\infty$ limit distribution, $\Phi_0(y)$, can be carried out with an excellent precision. 
The aim of this section is to present 
the steps of the calculation of the leading and next-to-leading FS corrections to the 
limit distribution.

We start by expanding the logarithm in the exponential on the r.h.s. of Eq.(\ref{analytic_Pm})
which allows rewriting the equation in terms of the coefficients $g_n$ defined by Eq.\eqref{gndef}. After rescaling the 
integration variable by $\sqrt{g_2/2}$, we obtain $\Phi_L(y)$ as the following 
Fourier integral
\begin{equation} \label{PDF_gn}
  \Phi_L(y) = \int\displaylimits_{-\infty}^{\infty} \frac{dk}{2\pi} 
  \exp{\left\lbrace iky 
      - \frac{1}{2}k^2 + F_L(k) \right\rbrace},
\end{equation}
where we have defined:
\begin{equation} \label{PsiLdef}
  F_L(k) = \sum_{n \geq 3} \frac{g_n}{2n}\left( ik\sqrt{\frac{2}{g_2}}\right)^n.
\end{equation}
The $L$ dependence of the above sum is in the coefficients $g_n$
which have a finite non-zero thermodynamic limit $g_n(L\to\infty)=g_n^\infty$ for $n \geq 2$. Thus,
in order to compute the FS behavior of $\Phi_L(y)$, we shall have to determine the FS corrections 
to $g_n^\infty$
\begin{equation} \label{gn_corr}
  g_n = g_n^\infty + \delta g_n\,.
\end{equation}
Assuming that $\delta g_n$ is known, we can write $F_L(k)$ as
\begin{equation} \label{Psi_separation}
F_L(k) = F_0 (k) + \delta
F_1 (k) + \delta F_2 (k) 
\end{equation}
where  $F_0 (k)$ is the thermodynamic limit of $F_L(k)$:
\begin{equation} \label{Psi0}
  F_0 (k) = \sum_{n \geq 3} \frac{g_n^\infty}{2n} 
  \left( ik \sqrt{\frac{2}{g_2^\infty}} \right)^n 
\end{equation}
and the FS corrections, due to $\delta g_2$ and $\delta g_n$ for $n \ge 3$, 
are written separately
\begin{eqnarray} \label{PsiA}
  \delta F_1 (k) &=&  - \frac{\delta g_2}{2 g_2^\infty} 
  \sum_{n \geq 3} \frac{g_n^\infty}{2} 
  \left( ik \sqrt{\frac{2}{g_2^\infty}} \right)^n \,,
  \qquad \\
  \delta F_2 (k) &=& \sum_{n \geq 3} \frac{\delta g_n}{2n} 
  \left( ik \sqrt{\frac{2}{g_2^\infty}} \right)^n 
  \,.\label{PsiB}
\end{eqnarray}
The separation of the $\delta g_2$ and the $\delta g_{n\ge 3}$ contributions 
is partly motivated by their $L\to\infty$ asymptotic behaviors. It will be shown in the 
Appendices that $\delta g_2\sim \ln L /L^2$ while for $n \ge 3$, $\delta g_{n}\sim 1/L^2$. 
Thus the leading correction comes from $\delta F_1 (k)$ and the sum in Eq.\eqref{PsiA} determines
the shape (the functional form) of the leading correction. As it turns out, the same shape 
correction can be easily separated from the $\delta g_{n}$ contributions for $n \ge 3$. 
Indeed, for large $n$,  one has $\delta g_n\sim n g_n^\infty$, and one can write
(cf. Eqs.\eqref{expan_dgn}, \eqref{gninfA} and \eqref{dgnxA})
\begin{equation} \label{dgncn}
  \delta g_n = \frac{\pi^2}{3 L^2} n g_n^\infty + \frac{2\pi}{3 L^2} c_n \hspace{0.2cm},
\end{equation}
where the second term is suppressed relative to the first one by 
a factor $4^{-n}$. Substituting the above split of $\delta g_n$ into Eq.\eqref{PsiB}, 
one can see the emergence of the same sum as in Eq.\eqref{PsiA}, and it allows us to write
\begin{eqnarray} \label{dPsik}
  \delta F_1(k) + \delta F_2(k) 
  = a_1(L) \Psi_1(k) + a_2(L) \Psi_2(k)
\end{eqnarray} 
where the $L$-dependent amplitudes are given by
\begin{equation} \label{a12}
  a_1(L) = \frac{\delta g_2(L)}{2 g_2^\infty} - \frac{\pi^2}{3 L^2} \,,
  \qquad
  a_2(L) = \frac{2\pi}{3 L^2}\,,
\end{equation}
while the corresponding $L$-independent functions by
\begin{eqnarray} \label{Psi12}
  \Psi_1(k) &=& -\sum_{n \geq 3} \frac{g_n^\infty}{2} 
  \left( ik \sqrt{\frac{2}{g_2^\infty}} \right)^n
  = - k \frac{d}{d k} F_0(k) \,,
\qquad \nonumber \\
  \Psi_2(k) &=& \sum_{n \geq 3} \frac{c_n}{2n} 
  \left( ik \sqrt{\frac{2}{g_2^\infty}} \right)^n \,.
\end{eqnarray} 
Integral representations for $F_0(k)$ and $\Psi_2(k)$ are given
in Eqs.\eqref{Psi0_int} and \eqref{Psi2_int}, respectively, in Appendix B.

Inserting Eq.\eqref{dPsik} into Eq.\eqref{PDF_gn},
we obtain the PDF of the $2d$ $XY$-model 
at zero-temperature, including its leading FS corrections
\begin{equation} \label{FSS_Pm}
  \Phi_L(y) = \Phi_0 (y) + a_1(L) \, \Phi_1(y) + a_2(L) \, \Phi_2(y) \hspace{0.2cm}.
\end{equation}
Here 
\begin{equation}\label{Phi0}
  \Phi_0 (y) = \int\displaylimits_{-\infty}^{\infty} \frac{dk}{2\pi} 
  \exp{\left\lbrace iky - \frac{k^2}{2} + F_0(k) \right\rbrace},
\end{equation}
and
\begin{equation}  \label{Phi12}
  \Phi_{1,2}(y) = \hspace{-0.1cm}\int\displaylimits_{-\infty}^{\infty} \hspace{-0.05cm} \frac{dk}{2\pi} 
  \exp{\left\lbrace iky - \frac{k^2}{2} + F_0(k) 
    \right\rbrace}\Psi_{1,2} (k) \hspace{0.1cm}.
\end{equation}

Since $\Psi_1(k) = -k F_0'(k)$, one can evaluate $\Phi_1(y)$ 
through $\Phi_0(y)$ by carrying out the appropriate 
integrations by part on the r.h.s. of Eq.\eqref{Phi12}. The outcome is one of the 
main results of our work, namely a simple expression is obtained for the leading shape 
correction in terms of the limit distribution (as quoted in Eq.\eqref{Phi_1}):
\begin{equation}  \label{Phi1y}
  \Phi_1 (y) = \left[\, y \,\Phi_0 (y) 
    + \Phi^\prime_0 (y) \right]^\prime.
\end{equation} 
The second shape correction  $\Phi_2(y)$ can not be related to $\Phi_0 (y)$ 
in such a simple manner but can be readily evaluated using Eqs.\eqref{Phi12} and \eqref{Psi2_int} of the Appendix \ref{Appendix_B}. 

\begin{figure}[H]
  \centering \hspace{0.5cm} \includegraphics[width=
  1.0\linewidth]{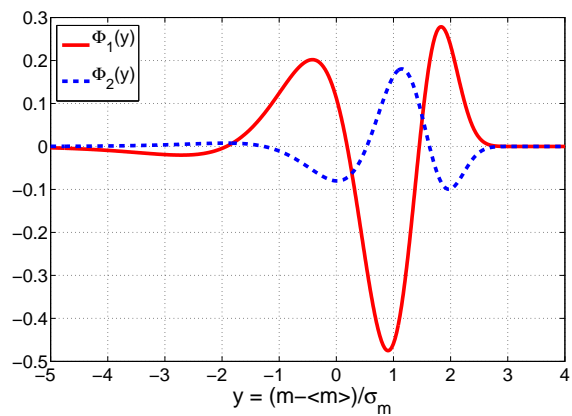} \vspace{-0.8cm}
  \caption{\textit{Behavior of the scale independent shape correction functions 
    $\Phi_1(y)$ and $\Phi_2(y)$.}}
  \label{FSS_Phi1_Phi2_fig}
\end{figure}

The functions $\Phi_1(y)$ and $\Phi_2(y)$ are displayed in 
Fig.\ref{FSS_Phi1_Phi2_fig}.
A general property of these functions is that their 0$^{th}$, 1$^{st}$ and 
2$^{nd}$ moments are zero. This follows 
from their definition as being corrections to a centered and normalized probability distribution, 
and can be explicitly verified using their definitions, e.g. Eq.\eqref{Phi12}.

In order to determine the amplitudes in front of the shape corrections,  
we have to calculate the FS behavior of $\delta g_n$ for $n\ge 2$. 
This problem is addressed in details, using two distinct methods, in the 
Appendices \ref{Appendix_A} and \ref{Appendix_B}. It is found there that the
asymptotic $L$-dependence of $\delta g_2$ has the form 
$(C_1 \ln L + C_2)/L^2$ (cf. Eq.\eqref{log_corr_g2})
while $\delta g_n$ behaves as $\propto 1/L^2$ for $n>2$
(cf. Eqs.\eqref{corr_gn}, \eqref{gninfA} and \eqref{dgnxA}). 
An important result of these calculations is the amplitude of $\Phi_1(y)$ given by
\begin{equation}\label{a_1}
  a_1(L) = \frac{\alpha \ln L + \gamma}{L^2} 
  = \frac{\alpha}{L^2} \ln\frac{L}{L_0}
\end{equation}
where the coefficient $\alpha$ is obtained analytically 
$\alpha=3\pi/(4G)= 2.5723613476$ with $G$ being the Catalan constant, while 
$\gamma = -2.8025632653$ and $L_0=2.972759081$ are determined numerically.

\begin{figure}[htb]
  \centering \vspace{-0.3cm} \hspace{0.5cm} \includegraphics[width=
  1.05\linewidth]{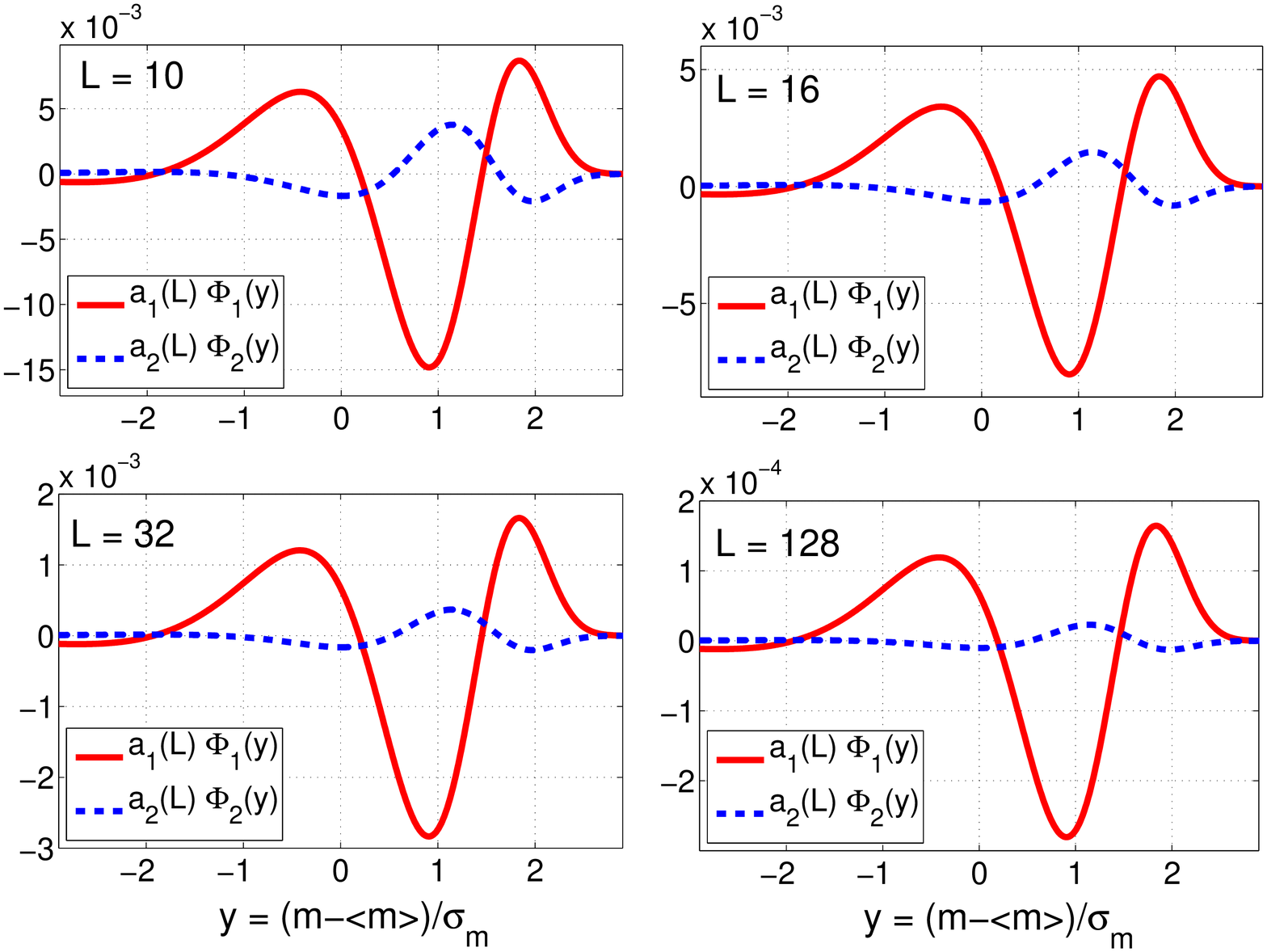} \vspace{-0.6cm}
  \caption{\textit{Comparison of the two leading correction terms $a_1(L) \Phi_1(y)$ and 
    $a_2(L) \Phi_2(y)$ for $L = 10, 16, 32 $ and 128.}}
  \label{a1Phi1_vs_a2Phi2_fig}
\end{figure}

The amplitude $a_1(L)$ is the second main result of our work since the 
FS corrections are dominated by the $a_1(L)\Phi_1(y)$ term. 
Indeed, $a_2(L)/a_1(L)$ is small, $0.67$, already for $L=10$ and it decreases with 
increasing $L$. Furthermore, evaluating $\Phi_2(y)$ numerically, one finds that apart from the 
neighborhood of the zeros of $\Phi_1(y)$, the inequality $a_1(L)\Phi_1(y)\gg a_2(L)\Phi_2(y)$ 
holds for $L> 10$ as seen in Fig.\ref{a1Phi1_vs_a2Phi2_fig}.

\begin{figure}[htb]
  \centering \vspace{-0.cm} \hspace{0.5cm} \includegraphics[width=
  1.0\linewidth]{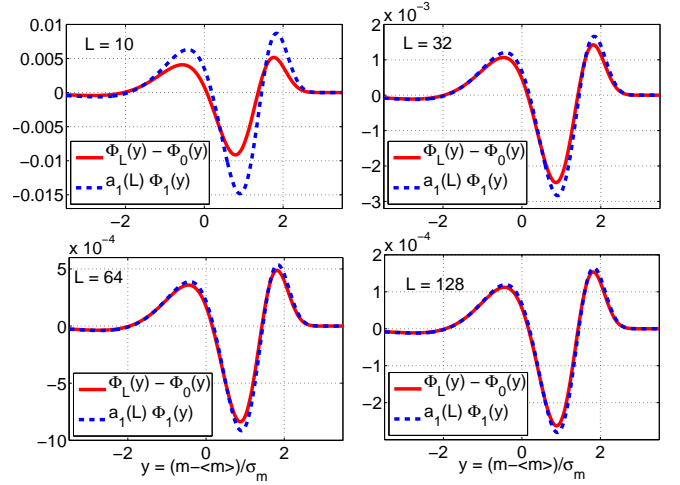} \vspace{-0.4cm}
  \caption{\textit{Comparison of the exact FS correction $\Phi_L(y)-\Phi_0(y)$  
  with the leading term $a_1(L) \Phi_1(L)$ given by \eqref{Phi1y} and \eqref{a_1}. 
  Apart from the regions close to the 
  maxima and minima, good agreement can be observed already for small system sizes.}}
  \label{dPhiL_vs_a1Phi1fig}
\end{figure}

It should be mentioned that there is some freedom
in separating the two contributions $a_1(L)\Phi_1(y)$ and $a_2(L)\Phi_2(y)$
in Eq.\eqref{lead-c1}. One can replace 
$\Phi_2(y)$ by $\Phi_2(y)+ c \Phi_1(y)$ changing simultaneously 
$\gamma$ to $\gamma-c\gamma'$ in $a_1(L)$, Eq.\eqref{a12}.
Our choice of separating the leading large-$n$ asymptote of $\delta g_n$ 
in Eqs.\eqref{dgncn} and \eqref{Psi12} leads naturally to a function proportional 
to $\Phi_1(y)$ and also results in a remnant that is small already for 
small values of $L$. This choice is convenient, since it allows to write the 
FS correction in a compact form [see Eq.\eqref{nextlead-c2}] with a very good accuracy 
as demonstrated in Fig.\ref{dPhiL_vs_a1Phi1fig}.

It is remarkable that the dominant correction term $a_1(L)\Phi_1(y)$ 
also emerges from a simple assumption about the PDF written in its original variable. 
Namely, if we assume that one can write $P_L(m) = P_0(m) + \epsilon(L) P''_0(m)$
with $\epsilon \to 0$ for $L\to\infty$, we find that
$\epsilon(L)=a_1(L)\, \sigma_m^2$. Using then the 
scaled variable $y$, the expression $\epsilon(L) P''_0(m)$ becomes $a_1(L)\Phi_1(y)$.

\begin{figure}[H]
  \centering \vspace{-0.4cm} \hspace{0.5cm} \includegraphics[width=
  1.05\linewidth]{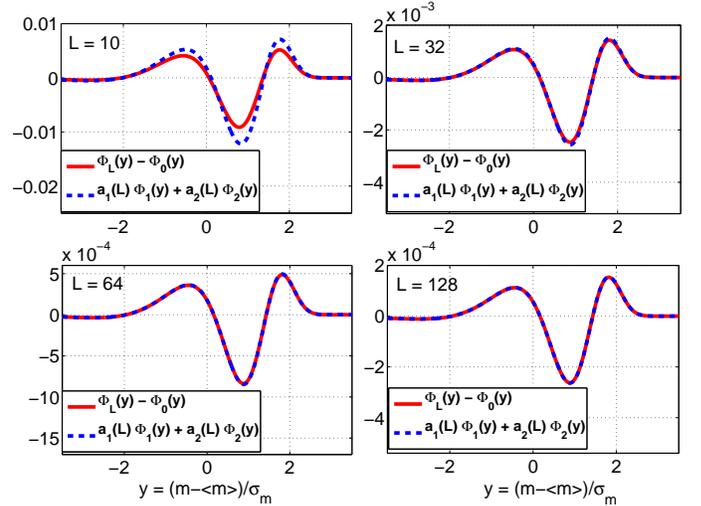} \vspace{-0.4cm}
  \caption{\textit{Comparison of the exact FS correction $\Phi_L(y)-\Phi_0(y)$  
  with the leading terms $a_1(L) \Phi_1(L)+a_2(L)\Phi_2(y)$ for lattice 
  sizes $L = 10, 32, 64$ and 128.}}
  \label{FSS_corr_analys_fig}
\end{figure}

As discussed above, there are other choices for separating a contribution proportional 
to $\Phi_1(y)$. It should be clear, however, that the freedom is irrelevant when the 
sum of the two contributions $a_1(L)\Phi_1(y)+a_2(L)\Phi_2(y)$ is used. As expected, 
and as can be seen 
in Fig.\ref{FSS_corr_analys_fig}, the convergence is fastest when the sum of both 
corrections are used.

% *****************************************************************************
\section{Monte Carlo simulations}
\label{MC_sect}
% *****************************************************************************

Here, we briefly demonstrate that the calculated FS corrections can be 
observed in simulations. It is clear that increasing the system size and improving 
the statistics by increasing the number of Monte Carlo samples, the leading FS corrections,
as seen in Fig.\ref{FSS_corr_analys_fig}, will emerge from the analysis. 
The question is whether the leading corrections 
we calculated could be seen already in small systems with reasonable simulation effort.

We have thus performed MC simulations on a $2d$ $XY$-model of size $L=10$ and
computed $\Phi_L (y)$.  
Since our analytic results Eqs.\eqref{FSS_Pm}-\eqref{Phi12} pertain to the $T=0$
limit of the system,  
we made simulations deep in the low $T$ region ($T = 0.04$ and $0.02$) using
the  over-relaxation  
Metropolis (ORM) algorithm \cite{over_relax_MC}.
In addition, simulations of the $T=0$ limit itself could also be carried out
since there the  
spin-wave approximation applies which yields independent modes with 
Gaussian action whose simulation is straightforward \cite{gaussian}.

\begin{figure}[H]
  \centering \vspace{-0.1cm} \hspace{0.5cm} \includegraphics[width=
  1.0\linewidth]{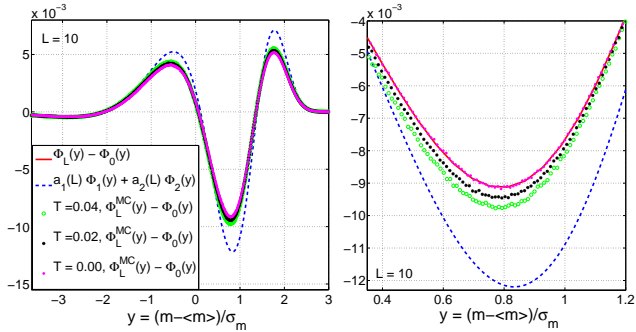} \vspace{-0.2cm}
  \caption{\textit{Comparison of the exact FS correction $\Phi_L(y)-\Phi_0(y)$ (red line) 
  and the leading terms $a_1(L) \Phi_1(L)+a_2(L)\Phi_2(y)$ (blue dashed line) with 
  the MC simulations for lattice size $L = 10$.
  The region close to the minimum is magnified in the right panel. It shows that 
  the ORM results approach to $\Phi_L(y)-\Phi_0(y)$ as $T \rightarrow 0$. } 
}
  \label{FSS_corr_analys_MC_fig}
\end{figure}

In the simulation using ORM, we typically measured observables
  using  $100 \times 10^9$ sweeps, and the errors were estimated by using a
  binning method.  
On Fig.\ref{FSS_corr_analys_MC_fig}, the red line shows the  
FS corrections to $\Phi_0 (y)$ displayed as the difference 
$\Phi_{L=10} (y) - \Phi_0 (y)$ computed from the integral representation 
of Eq.\eqref{analytic_Pm}. It is compared with data obtained by ORM at very low 
temperatures $T = 0.04$ (green circles) and $T = 0.02$ (asterisk markers). 
The magenta points represent the results from the simulation of the Gaussian 
action at $T = 0$.
The statistical errors on all the data points displayed are smaller than
the point size, which is not surprising for the quoted large number
of MC sweeps. From the actual statistical error $\sim 5\times 10^{-5}$
one can find that a relatively small number, $3\times 10^8$ sweeps 
are enough to reach an accuracy $10^{-3}$, one tenth of the maximal FS correction.

As one can see, the simulation temperatures used are small enough for the
temperature corrections to be small compared to the FS corrections at $L=10$.
It can also be seen that the the sum of the two leading FS correction terms,
$a_1 (L) \Phi_1(y) + a_2 (L) \Phi_2(y)$ (blue dashed line)
is quite close to the exact result. 
Of course, $L=10$ is a rather small system to expect full agreement 
of the calculated leading FS corrections with the full FS correction. 
Looking at Fig.\ref{FSS_corr_analys_fig}, however, one notes that increasing 
the size of the system by only a factor three would yield a complete
domination of the leading FS corrections over the higher order FS corrections.

Thus, we conclude that observing the leading FS corrections is feasible in
relatively small systems at relatively small computational coast. 
 Based on this observation, we expect that a meaningful analysis of FS 
corrections in experimental $XY$ systems is also possible.

% *****************************************************************************
\section{Conclusions}
\label{Conclus_sect}
% *****************************************************************************

We have computed the leading FS corrections to the PDF of the magnetization 
of the $2d$ $XY$-model at zero temperature. Two
scale-independent functions $\Phi_1(y)$ and $\Phi_2(y)$ were found with their amplitudes 
behaving with system size as $a_1(L)\sim \alpha \ln(L/L_0)/L^2$ and $a_2(L)\sim 1/L^2$.
The function $\Phi_1(y)$ can be expressed through 
the limit distribution $\Phi_0(y)$ and its low-order derivatives. This makes 
it a candidate for identifying universality features hidden in FS corrections.

The leading and next to leading corrections were found to describe the FS behavior 
very accurately already for small system size. Thus, as our MC simulations demonstrated,
the observation of the calculated FS corrections is possible in model systems. We expect
that their experimental observation may also be feasible.

% *****************************************************************************
\section*{Acknowledgments}
% *****************************************************************************

This work was partially supported by Dicyt-USACH Grant No. 041531PA and
PAI-CONICYT 79140064, and by the Hungarian Research Fund (OTKA NK100296). 
G. P. would like to thank the Institute for Theoretical Physics at E\"otv\"os 
University, Budapest, for the invitation in the summer of 2014, where
 this collaboration started and also thanks to their members for the kind hospitality.

% *****************************************************************************
\begin{appendix}
% *****************************************************************************
\section{Finite-size corrections to $\delta g_n$}
\label{Appendix_A}
% *****************************************************************************

We begin by deriving the large-$L$ asymptotic expansion for $g_2$ defined in
Eq.(\ref{gndef}). It is convenient to write this
equation in the form:
\begin{eqnarray} \label{g2_asymp_exp}
g_2 &=& \frac{1}{(2\pi)^4} \sum_{l,m}
    \hspace{-0.1cm}' \left[ l^2 + m^2 
      -\frac{1}{12}\left(\frac{2 \pi}{L} \right)^2 \hspace{-0.2cm}(l^4 + m^4) 
      + \cdots \right]^{-2} \nonumber\\
    &=& g_2^\infty + \delta g_2
\end{eqnarray}
where the sum goes from $l,m = -L/2 + 1$, to $L/2$ and prime means that 
the $l = m = 0$ term is left out. The asymptotic value $g_2^\infty$ is given by
\begin{eqnarray} \label{g2_inf} 
g_2^\infty &=& \frac{1}{(2\pi)^4}
  \sum_{l,m=-\infty}^\infty \hspace{-0.3cm}'\hspace{0.3cm} \frac{1}{(l^2 +
    m^2)^2} = G/(24 \pi^2) \nonumber \\
    &\approx& 0.0038669
\end{eqnarray} 
where $G$ is the Catalan's constant.  

In the FS correction $\delta g_2$, the sum giving the coefficient of
$1/L^2$ diverges logarithmically with $L$. This leading term is obtained as
\begin{eqnarray} \label{log_corr_g2}
    \delta g_2 &\sim& \frac{1}{6(2 \pi)^2 L^2} \sum_{l^2 + m^2 < L^2/4} 
    \hspace{-0.6cm}'\hspace{0.5cm} \frac{l^4 + m^4}{(l^2 + m^2)^3} + \cdots \nonumber\\
    &=& \frac{1}{6(2 \pi)^2 L^2} 
    \int\displaylimits_{1/L} ^{1/2} \frac{dr}{r} 
    \int\displaylimits_0^{2\pi} d\phi (\sin^4{\phi} + \cos^4 {\phi}) + \cdots \nonumber\\
    &=& \frac{1}{16 \pi}\frac{\ln{L}}{L^2} + \mathcal{O}(L^{-2}).
\end{eqnarray}
It is worthwhile to mention that the $\ln L/L^2$ decay of $\delta g_2$ is consistent with the
logarithmic FS corrections of some related quantities reported in \cite{log_corrections}.  

The above expression for $\delta g_2$ can be generalized to perform a high
precision fit of the form:
\begin{equation} \label{g2_fit}
  g_2(L) = g_2^\infty + \frac{1}{L^2}\left( \frac{1}{16\pi}\ln L + \gamma_2
  \right) \,,
\end{equation}
to $g_2(L)$ computed numerically using Eq.\eqref{gndef} for a large
range of $L$ values \cite{footnote2}.

\begin{figure}[H]
  \centering \vspace{-0.4cm} \hspace{-0.5cm} \includegraphics[width=
  1.05\linewidth]{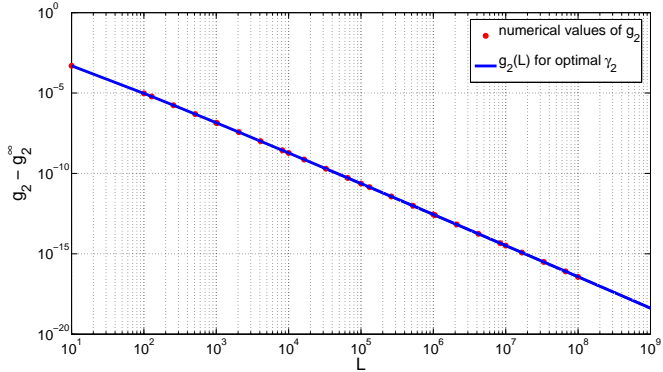} \vspace{-0.5cm}
  \caption{\textit{Behavior of $g_2(L)$ as a function of lattice size
    $L$ up to $L = 10^9$ in logarithmic scale. The red dots correspond to the
    direct numerical evaluation of Eq.(\ref{gndef}) and the blue line
    corresponds to the analytic expression of Eq.(\ref{g2_fit}) for 
    $\gamma_2 = 0.003768763799$.} }
  \label{g2_fit_fig}
\end{figure}

In Fig.\ref{g2_fit_fig} we have plotted the numerically computed $g_2(L)$, 
as well as the FS scaling
expression given by Eq.(\ref{g2_fit}). The value of the parameter 
$\gamma_2 = 0.003768763799$ was obtained from a high precision fit 
over system sizes up to $L=10^9$.
In Appendix \ref{Appendix_B} we use a more sophisticated method to obtain 
an integral representation for $\gamma_2$, and found a complete
agreement with the value cited above.
Similarly to Eq.\eqref{g2_inf} the asymptotic value $g_n^\infty$ is given by
\begin{eqnarray} \label{gn_infty}
  g_n^\infty &=& \frac{1}{(2\pi)^{2n}}
  \sum_{l,m= -\infty}^\infty \hspace{-0.3cm}'\hspace{0.2cm} 
  \frac{1}{(l^2 + m^2)^n} \nonumber \\ 
  &=& \frac{4}{(2\pi)^{2n}}  \zeta (n) \beta (n)
\end{eqnarray}
where $\zeta(n)$ and $\beta(n)$ are Riemann's zeta function and Dirichlet's
beta function, respectively \cite{math_func}.

For $n\ge 3$ the sum appearing in the $1/L^2$ correction term $\delta g_n$
converges for $L\to\infty$ hence one can extend the summation to $\pm\infty$
(up to an error decreasing faster than $1/L^2$). One has then
\begin{eqnarray} \label{expan_dgn}
    \frac{\delta g_n}{n g_n^\infty} &=& \frac{ (2\pi)^2}{12 L^2} 
    \frac{\sum_{l,m}\hspace{-0.5cm}' \hspace{0.4cm} 
      (l^4 + m^4)(l^2 + m^2)^{-(n+1)}} {\sum_{l,m} \hspace{-0.5cm}' 
      \hspace{0.4cm} (l^2 + m^2)^{-n}} \nonumber \\
%    &=& \frac{\pi^2}{3L^2} \left[1 +
%      \frac{\sum_{l,m}\hspace{-0.5cm}' \hspace{0.4cm}
%        (l^4 - l^2 +m^4 - m^2)(l^2 + m^2)^{-(n+1)}}{\sum_{l,m}
%        \hspace{-0.5cm}'\hspace{0.4cm}(l^2 + m^2)^{-n}}  \right] \\
    &=& \frac{\pi^2}{3L^2} \left( 1 + \frac{U_n}{V_n}\right)
\end{eqnarray}
where
\begin{eqnarray}
  U_n &=& \sum_{l,m}\hspace{-0.05cm}' \hspace{0.2cm}
  \frac{l^4 -l^2 +m^4 -m^2}{(l^2 + m^2)^{n+1}} \, \qquad  \nonumber \\
  V_n &=& \sum_{l,m}\hspace{-0.05cm}' \hspace{0.2cm} 
  (l^2 + m^2)^{-n}= 4\zeta(n) \beta(n) \hspace{0.2cm}.
\end{eqnarray}
For large $n$ the dominant terms in these sums come from smallest $|l|$, $|m|$
with non-vanishing contributions.  The numerator in $U_n$ vanishes for the two
shells $(l=\pm 1, m=0)$, $(l=0, m=\pm 1)$ and $(l=\pm 1, m=\pm 1)$. The
leading term for large $n$ is coming from $(l=\pm 2, m=0)$, $(l=0, m=\pm 2)$
and is given by $12\cdot 4^{-n}$.
Since $V_n = 4 (1+ 2^{-n}+\ldots)$ one finds that $U_n/V_n \sim 3 \cdot
4^{-n}$.  This is a small correction -- even for $n=3$ it is just $\approx
0.047$.
Hence we have, as stated in Eq.\eqref{dgncn}
\begin{equation}  \label{corr_gn}
  \frac{\delta g_n }{n g_n^\infty} = \frac{\pi^2}{3L^2}
  + \frac{1}{L^2} \order{4^{-n}}\,.
  \qquad n\ge 3
\end{equation}
%

% *****************************************************************************
\section{Integral representation for leading
finite-size corrections}
\label{Appendix_B}
% *****************************************************************************
For calculating finite-volume sums for a cubic box of size $L$ in $d$
dimensions in the continuum, like 
$\sum_\mathbf{k}' 1/(\mathbf{k}^2)^{n}$, where 
$\mathbf{k}^2= \sum_{i=1}^d (2\pi n_i /L)^2$, it is useful to introduce
the function $S(x)$ (see e.g. \cite{Has90}),   
(related to Jacobi's theta function) defined as
\begin{equation} \label{Sdef}
	S(x) = \sum_{n=-\infty}^{\infty}
	\exp\left(-\pi x n^2\right).
\end{equation}
It satisfies the relation
\begin{equation} \label{Srel}
  S(x) = \frac{1}{\sqrt{x}}S\left(\frac{1}{x}\right)
\end{equation}
which allows to calculate $S(x)$ very precisely by taking only a few terms 
in the sum, both for $x<1$ and $x>1$.
Note that $S(x)= x^{-1/2}(1+2\exp(-\pi/x)+\ldots)$ for $x\to 0$, 
while $S(x)=1 + 2 \exp(-\pi x) + \ldots$ for large $x$.

As an illustration, it is easy to show that $g_n^\infty$ given by 
Eq.\eqref{gn_infty} has an integral representation
\begin{equation} \label{gninf}
  g_n^\infty = \frac{1}{(4\pi)^n\Gamma(n)} 
  \int\displaylimits_0^\infty \mathrm{d}x \, x^{n-1}
  \left( S^2(x)-1\right) \,.
\end{equation}
The leading term for $n\to\infty$ is given by the
large-$x$ behavior of the integrand. Separating it, one obtains an expression
\begin{eqnarray} \label{gninfA}
  g_n^\infty = \frac{4}{(2\pi)^{2n}} +\hspace{-0.4cm}&& 
 \frac{1}{(4\pi)^n\Gamma(n)} \times \nonumber \\
 &&\int\displaylimits_0^\infty \mathrm{d}x \, x^{n-1}\hspace{-0.1cm}
  \left( S^2(x)-1 -4 \mre^{-\pi x} \right) 
\end{eqnarray} 
which can be evaluated and shown to be in agreement with Eq.\eqref{gn_infty}. 
With the help of this one can perform the summation in Eq.\eqref{Psi0}
yielding
\begin{eqnarray} \label{Psi0_int}
F_0(k) = \int_0^\infty \hspace{-0.2cm}&&\frac{\mrd u}{2u} 
\left[S^2\left(u w\right)-1\right]\times \nonumber \\
&&\left\lbrace
\mre^{iku}
 - \left(1+iku-\frac12 k^2 u^2\right)
\right\rbrace
\end{eqnarray} 
where $w=4\pi \sqrt{g_2^\infty/2}$.
The Fourier transformation appearing here can be performed efficiently
by a fast Fourier transform (FFT).

This technique can be generalized to finite-volume lattice sums
by introducing \cite{Nie15b}
\begin{eqnarray} \label{QLdef}
  Q_L(z) &=& \frac{1}{L}	\sum_{l=0}^{L-1} \exp\left(-z \hat{k}_l^2\right) \nonumber \\
	&=& \phi_0(z) + 2 \sum_{m=1}^\infty \phi_{mL}(z)
\end{eqnarray} 
where $\hat{k}_l^2= 2(1-\cos(2\pi l/L))$,  $l=0,\ldots,L-1$, and 
\begin{equation}
\phi_n(z) = \mathrm{e}^{-2z} I_n(2z)
\end{equation}
with $I_n(z)$ being the modified Bessel function.
For large $z$ one has $\phi_0(z)\sim 1/\sqrt{4\pi z}$.

For fixed $z$ with increasing $L$ the function
$Q_L(z)$ approaches $\phi_0(z)$ exponentially fast.
The approach becomes slower with increasing $z$,
but even when the argument increases slower than $L^2$ one still has
\begin{equation}
\lim_{L\to\infty} \left(Q_L(c L^\alpha) - \phi_0(c L^\alpha)\right)
= 0 \qquad \text{for } \alpha < 2
\end{equation}
with the difference decreasing faster than any inverse
power of $L$. 
This is not true for $z\propto L^2$, and for 
this case one obtains another scaling function.
Rescaling $Q_L(z)$, we introduce the lattice counterpart \cite{Nie15b}
of $S(x)$ by
\begin{equation} \label{SLdef}
  S_L(x) = L Q_L\left( \frac{x L^2}{4\pi}\right)	
\end{equation}
By expanding Eq.\eqref{QLdef} for large $L$ one finds the asymptotic expansion
\begin{equation} \label{SL_as}
S_L(x) =  S(x) + \frac{\pi}{3L^2} x S''(x)
 + \mathcal{O}\left(\frac{1}{x^2 L^4} \right).
\end{equation}
As the error term indicates, the approach to $L=\infty$
is not uniform in $x$.

Using $S_L(x)$ one has for $g_n=g_n(L)$ two integral representations
\begin{eqnarray} \label{gn}
    g_n &=& \frac{L^{-2n+2}}{\Gamma(n)}\int_0^\infty\mrd z\, z^{n-1}
    \left[Q_L^2(z)-\frac{1}{L^2} \right] \nonumber \\
    &=& \frac{1}{(4\pi)^n \Gamma(n)}
    \int_{0}^\infty \mrd x\, x^{n-1} 
    \left[S_L^2(x)-1\right].
\end{eqnarray} 
We outline below the calculation of $\delta g_2$ to $\order{1/L^2}$.
Due to the non-uniform convergence for $L\to\infty$
it is useful to split the integration region and write
\begin{eqnarray} \label{g2L}
g_2(L) = 
\frac{1}{L^2} \int_0^{z_0} \hspace{-0.4cm}&& \mrd z \, z \left[
Q^2_L(z)-\frac{1}{L^2}\right]
+ \nonumber \\
&&\frac{1}{(4\pi)^2}\int_{x_0}^\infty \mrd x \, x \left[ S_L^2(x)-1\right]
\end{eqnarray}  
where $x_0=4\pi z_0/L^2$. Choosing $z_0=z_0(L)= c L^{2-\epsilon}$ 
with some fixed small $\epsilon>0$ in the first term one could replace 
$Q_L(z)$ by $\phi_0(z)$
up to exponentially small corrections.
Similarly, in the second integral one can use the expansion Eq.\eqref{SL_as}.
Note that $z_0(L)\to\infty$ and $x_0(L)\to 0$  for $L\to\infty$.
Using Eq.\eqref{SL_as} and neglecting terms vanishing faster than $1/L^2$
one obtains
\begin{equation} \label{dg2L}
  \delta g_2 = 
  \frac{1}{L^2} \int_0^{z_0} \mrd z \, z \phi_0^2(z)
  +\frac{2\pi}{3 L^2 (4\pi)^2}\int_{x_0}^\infty \mrd x \, x^2 S(x) S''(x)
\end{equation}
Separating the asymptotic behavior of the integrands for large $z$,
and small $x$, respectively, one obtains the logarithmic contribution
$1/(32\pi) \ln(z_0/x_0)= 1/(16\pi) \ln L$, and in the remaining terms
one can make the substitutions $z_0=\infty$ and $x_0=0$.
Evaluating the corresponding integrals one reproduces the fit result Eq.\eqref{g2_fit} 
to all digits (cf. Fig.\ref{g2_fit_fig}).

The leading correction of $\delta g_n$ for $n>2$ is simpler and given by
the convergent integral
\begin{equation} \label{dgnx}
\delta g_n =  \frac{2\pi}{3 L^2}
\frac{1}{(4\pi)^n \Gamma(n) }
\int_{0}^\infty \mrd x\, x^{n} S(x) S''(x) \,, \quad n\ge 3
\end{equation}
Separating the large-$x$ term of the integrand one obtains
 \begin{eqnarray}  \label{dgnxA}
 \delta g_n =\hspace{-0.2cm}&& \frac{\pi^2}{3 L^2} \frac{4n}{(2\pi)^{2n}}
 +\frac{2\pi}{3 L^2}
 \frac{1}{(4\pi)^n \Gamma(n) } \times \nonumber \\
 &&\int_{0}^\infty \mrd x\, x^{n} 
 \left[S(x) S''(x) - 2\pi^2\mre^{-\pi x} \right]
 \end{eqnarray} 
where $n\ge 3$. The leading term has the same form as for $n g_n^\infty$
(cf. Eq.\eqref{gninfA}). 
Subtracting this way the leading term one can define $c_n$ by
\begin{equation} \label{dgnxB}
  \delta g_n = \frac{\pi^2}{3 L^2} n g_n^\infty + \frac{2\pi}{3 L^2} c_n
\end{equation}
where for $n\ge 3$:
\begin{equation} \label{cn_int}
  c_n =  \frac{1}{(4\pi)^n \Gamma(n) }
  \int\displaylimits_{0}^\infty \mrd x\, x^{n} S(x) 
  \left(S''(x) + \pi S'(x)\right).
\end{equation}

For large $n$ one has $c_n \sim 6\pi n (4\pi)^{-2n}$, i.e. it is suppressed
by a factor of $4^{-n}$ compared to $n g_n^\infty$.

Inserting Eq.\eqref{cn_int} into Eq.\eqref{Psi12} we obtain 
an integral representation for $\Psi_2(k)$,
\begin{eqnarray} \label{Psi2_int}
  \Psi_2(k) =
  \frac{w}{2} \hspace{-0.2cm}&&\int\displaylimits_0^\infty \mrd u \, S( u w )
  \left\lbrace S''(u w) + \pi S'(uw)\right\rbrace\times  \nonumber \\
  && \quad \left[\mre^{iku} - \left(1+iku-\frac12 k^2 u^2\right)\right].
\end{eqnarray} 
Using FFT one can evaluate this and finally  
the corresponding correction $\Phi_2(y)$ to the PDF.
\end{appendix}
%-----------------------------------------------------------------------------


\begin{thebibliography}{99}

	% 1
\bibitem{Fisher} M. E. Fisher in Critical Phenomena, Proceedings of the 1970
  International School of Physics Enrico Fermi, Course 51, Edited by
  M. S. Green (Academic, New York, 1971).

	% 2
\bibitem{CardyFSS} J. L. Cardy, Finite Size Scaling, (North-Holland, Amsterdam,
  1988).

	% 3
\bibitem{FSS_Num_stat} \textit{Finite Size Scaling and Numerical Simulation of
    Statistical Physic}s, edited by V. Privman (World Scientific, Singapore,
  1990).

	% 4
\bibitem{Bruce} A. D. Bruce, J. Phys. C \textbf{14}, 3667 (1981).

	% 5
\bibitem{Binder} K. Binder, Z. Phys. B \textbf{43}, 119 (1981).

	% 6
\bibitem{Nicolaides} D. Nicolaides, A. D. Bruce, J. Phys. A \textbf{21}, 233 (1988).
	
	% 7
\bibitem{Ex_wide_applic} Examples of wide-ranging applications are:
  A. D. Bruce and N. B. Wilding, Phys. Rev. Lett. \textbf{68}, 193 (1992) (liquid-gas
  transition), D. Nicolaides and R. Ewans Phys. Rev. Lett. \textbf{63}, 778 (1989)
  (wetting); N. B. Wilding and P. Niebala, Phys. Rev. E \textbf{53}, 926 (1996)
  (tricritical point); M. M\"{u}ller and N. B. Wilding, Phys. Rev. E \textbf{51}, 2079
  (1995) (polymers); S. L. A. de Queiroz and R. B. Stinchcomb, Phys. Rev. E
  \textbf{64}, 036117 (2001) (random-field Ising model); M. M. Tsypin,
  Phys. Rev. Lett. \textbf{73}, 2015 (1994) (field theory).

	% 8
\bibitem{Racz_et_al} G. Foltin, K. Oerding, Z. R\'{a}cz, R. L. Workman and
  R. K. P. Zia, Phys. Rev. E \textbf{50}, R639 (1994); Z. R\'{a}cz and M. Plischke,
  Phys. Rev. E \textbf{50}, 3530 (1994); E. Marinari, A. Pagnani, G. Parisi, and
  Z. R\'{a}cz, Phys. Rev. E \textbf{65}, 026136 (2002).

	% 9
\bibitem{Bramwell_nature} S. T. Bramwell, P. C. W. Holdsworth, J. -F. Pinton,
  Nature \textbf{396}, 552 (1998); R. Labb\'e, J. -F. Pinton, P. C. W. Holdsworth,
  Phys. Rev. E \textbf{60}, R2452 (1999).

	% 10
\bibitem{Aji_Goldenfeld} V. Aji and N. Goldenfeld, Phys. Rev. Lett. \textbf{86}, 1007
  (2001)

	% 11
\bibitem{Bramwell_prl} S. T. Bramwell et al., Phys. Rev. Lett. \textbf{84}, 3744
  (2000).

	% 12
\bibitem{Korniss_et_al} G. Korniss, Z. Toroczkai, M. A. Novotny, and
  P. A. Rikvold, Phys. Rev. Lett. \textbf{84}, 1351 (2000); S. Lubeck and P. C. Heger,
  Phys. Rev. Lett. \textbf{90}, 230601 (2003); T. Halpin-Healy and G. Palasantzas, EPL
  \textbf{105}, 50001 (2014).

	% 13
\bibitem{Kosterlitz_Thouless} J. M. Kosterlitz and D. J. Thouless, J. Phys. C
  \textbf{6}, 1181 (1973).

	% 14
\bibitem{Chaikin} P. M. Chaikin and T. C. Lubensky, Principles of Condensed
  Matter Physics (Cambridge University Press, Cambridge, 1996).

	% 15
\bibitem{Toth_prl} T. T\'{o}th-Katona and J. Gleeson, Phys. Rev. Lett. \textbf{91}
  264501 (2003); S. Joubaud, A. Petrosyan, S. Ciliberto, and N. B. Garnier,
  Phys. Rev. Lett. \textbf{100}, 180601 (2008).

	% 16
\bibitem{Bramwell_epl} S. T. Bramwell, T. Fennel, P. C. W. Holdsworth, and
  B. Portelli, EPL \textbf{57}, 310 (2002); K. Dahlstedt and H. Jensen, Physica A \textbf{348},
  596 (2005).

	% 17
\bibitem{Edwards_Wilkinson} S. F. Edwards and D. R. Wilkinson,
  Proc. R. Soc. London, Series A \textbf{381}, 17 (1982).
  
	% 18
\bibitem{Bramwell_PRE2001} S. T. Bramwell, J.-Y. Fortin, P. C. W. Holdsworth, S. Peysson, J.-F. 
Pinton, B. Portelli, and M. Sellito, Phys. Rev. E \textbf{63},
  041106 (2001).
  
	% 19
\bibitem{Bramwell_nat_phys2009} S. T. Bramwell, Nature Physics \textbf{5}, 444 (2009).
	
	% 20
\bibitem{Palma_2005} G. Mack, G. Palma and L. Vergara, Phys. Rev. E \textbf{72},
  026119 (2005).

	% 21
\bibitem{Palma_2006} G. Palma, Phys. Rev. E. \textbf{73}, 046130 (2006).
	
	% 22
\bibitem{Banks_2005} S. T. Banks and S. T. Bramwell, J. Phys. A \textbf{38}, 5603 (2005). 

	% 23  
\bibitem{Antal_et_al_prl} T. Antal, M. Droz, G. Gy\"{o}rgyi, and Z. R\'{a}cz,
  Phys. Rev. Lett. \textbf{87}, 240601 (2001); Phys. Rev. E. \textbf{65}, 046140 (2002).

	% 24
\bibitem{Archam_1998} P. Archambault, S. T. Bramwell, J.-Y. Fortin, P. C. W. Holdsworth, S. Peysson, J.-F. 
Pinton, J. Appl. Phys. \textbf{83}, 7234 (1998).

	% 25
\bibitem{Jona} G. Jona-Lasinio, Phys. Rep. \textbf{352}, 439 (2001).

	% 26
\bibitem{Cardy} J. Cardy, Scaling and Renormalization in Statistical
  Physics, Cambridge University Press (1996).

	% 27
\bibitem{over_relax_MC} M. Creutz, Phys. Rev. D \textbf{36}, 515 (1987); K. Kanki, 
D. Loison and K.D. Schotte, Eur. Phys. J. B \textbf{44}, 309?315 (2005)

	% 28
\bibitem{gaussian}
Integrating out the gaussian variables one obtains the analytic 
expression in Eq.\eqref{analytic_Pm}.

	% 27
\bibitem{log_corrections} R. Kenna and A.C. Irving, Phys. Lett B \textbf{351},
 273 (1995); W. Janke, Phys. Rev. B \textbf{55} (1997); S.G. Chung,
 Phys. Rev. B \textbf{60}, 16 (1999)


\bibitem{footnote2} Note that in the double sum appearing in \eqref{gndef} 
the summation over one of the integers can be done analytically, which makes 
possible to reach large values of $L$.



% 29
\bibitem{math_func} See, e.g., I. S. Gradstein and I. M. Ryshik, Table of
  Integral series and products, seventh edition, Elsevier Academic Press
  (2007)	


	% 30
\bibitem{Has90}
  P.~Hasenfratz and H.~Leutwyler,
  \emph{Nucl.\ Phys.\ B} {\bf 343}, 241 (1990).
	
	%31
\bibitem{Nie15b}
  F.~ Niedermayer and P.~Weisz,
  \emph{Massless sunset diagrams in finite asymmetric volumes},
  \emph{arXiv:1602.03159 [hep-lat].}
  

\end{thebibliography}
\end{document}